\documentclass[a4paper,12pt]{article}
\usepackage{graphicx}
\usepackage{amssymb,latexsym}
\usepackage[english]{babel}
\usepackage{graphicx}
\newcommand{\bec}{\begin{center}}
\newcommand{\ec}{\end{center}}
\newcommand{\bee}{\begin{equation}}
\newcommand{\ee}{\end{equation}}

\begin{document}
\large
\begin{titlepage}
\begin{center}
{\Large\bf {SEARCH FOR SUPERSYMMETRY\\ AT THE LARGE HADRON COLLIDER}}\\
{\bf T.V. Obikhod\\}
{\it Institute for Nuclear Research, Nat. Acad. of Sci. of Ukraine\\
47, Prosp. Nauky, Kyiv 03680, Ukraine\\}
{\bf V.V. Negliad\\}
{\it Taras Shevchenko National University of Kyiv\\
6, Prosp. Academician Glushkov, Kyiv 03127, Ukraine}\\
e-mail: obikhod@kinr.kiev.ua\\

\vspace*{11mm}
{\bf Abstract\\}
\end{center}
Search for supersymmetry is carried out in the framework of the
Minimal Supersymmetric Standard Model (MSSM).\,\,Using the software
programs SOFTSUSY and PROSPINO, the mass spectrum and the production
cross-sections of superpartners are calculated.\,\,The results
obtained are of importance for searching the new physics at the LHC.
\vspace*{3mm}
\end{titlepage}

\section{Introduction}

Problems in high-energy physics associated with unifying all
fundamental interactions in a single theory, the so-called Theory of
everything \cite{1}, bring about a necessity of creating high-energy
accelerators.\,\,On September 10, 2008, the Large Hadron Collider
(LHC)~-- the largest experimental installation in the world~-- was
officially started.\,\,The LHC can be useful not only in the study
of the Higgs mechanism responsible for the violation of the
electroweak interaction, but also in the verification of the theory
of superstrings and D-branes, because the latter predicts the
existence of such objects as

-- superpartners,

-- Kaluza--Klein particles,

-- dark matter candidates, and

-- microscopic black holes.

Our work aims at calculating the masses of superpartners at the LHC
in the framework of supersymmetry theory.\,\,The latter is one of
the most widespread theories beyond the Standard model (SM), because
it allows a number of SM problems to be solved.

The SM of particle physics is rather a successful theory for the
description of physical phenomena, which can be observed on modern
colliders.\,\,However, the theorists are sure that the SM will not
work at higher energy scales.\,\,Of course, the SM cannot be an
ultimate theory at very high energies, because it has to be modified
in order to include the gravitational interaction on Planck scales.
This problem is called the hierarchy problem \cite{2,3,4}.\,\,Even
if the SM gives the best description of the subatomic world, it does
not provide a complete picture of the Universe, since it does not
imply the unification mechanism for the strong and electroweak
interactions, on the one hand, and the gravitational interaction, on
the other hand \cite{1}. In addition, the radiation corrections to
the mass of a Higgs boson give rise to large discrepancies between
the experiment and the SM theory \cite{5}.\,\,A considerable amount
of the cold dark matter and the phenomenon of dark energy, which are
observed in experiment \cite{6}, go beyond the scope of SM
predictions.

Recent experimental data are bright markers of new physics beyond
the SM.\,\,In particular,

(i)~new measurements of the Higgs boson properties carried out by
ATLAS and CMS Collaborations \cite{7} point to the physics both in
the SM framework and beyond it;



(ii)~there are problems with the stability of electroweak vacuum
\cite{8};

(iii)~the total width of a recently discovered Higgs boson is 5.4
times larger than the corresponding SM value \cite{9}; and

(iv)~the {production cross-sections of} $W^{+}W^{-}$ particles obtained on
the LHC in the course of the proton-proton collisions at the
energies $\sqrt{s}=7$ and 8~TeV testify to a substantial deviation
from the SM \cite{10}.

While searching for the supersymmetry, we calculated the production
cross-sections of superpartners at the LHC at the energy
$\sqrt{s}=14~\mathrm{TeV}$ and determined the lower limits for the
superpartner masses.

\section{Calculations of Mass Spectra and Production\\ Cross-Sections of
Superpartners}

Supersymmetry supplementing fermions with bosons and {\it vice
versa} cannot be an exact symmetry in the Nature, because fermions
and bosons must be degenerate with respect to the mass.\,\,In order
to develop a realistic model of high-energy physics, the
supersymmetry has to be broken.\,\,The issue concerning a
supersymmetric interaction and the condition of its violation
associated with the inclusion of soft supersymmetry-breaking terms
into the interaction potential is well consistent in the framework
of the so-called Minimal Supersymmetric Standard Model~\cite{11}.

The MSSM is defined by the superpotential\vspace*{-1mm}
\[
W=h^e_{ij}L_iH_1\overline{E}_j+h^d_{ij}Q_iH_1\overline{D}_j+
h^u_{ij}Q_iH_2\overline{U}_j  +\mu H_1H_2
\]\vspace*{-5mm}

and the potential of a soft supersymmetry
breaking\vspace*{-1mm}
\[
V=m_1^2|H_1|^2+m_2^2|H_2|^2-m_{12}^2\left(\!\epsilon_{ij}H^i_1H^j_2+{\rm
h.c.}\!\right)+
\]\vspace*{-8mm}
\[
+\,M_{\tilde{Q}}^2\left[\tilde{t}^*_L\tilde{t}_L+\tilde{b}^*_L\tilde{b}_L
\right]+ M^2_{\tilde{U}}\tilde{t}^*_R\tilde{t}_R\,+
+\,M^2_{\tilde{D}}\tilde{b}^*_R\tilde{b}_R+ M^2_{\tilde{L}}
[\tilde{\nu}^*\tilde{\nu}+\tilde{\tau}^*_L\tilde{\tau}_L]\,+
\]\vspace*{-8mm}
\[
+\, M^2_{\tilde{E}}\tilde{\tau}^*_R\tilde{\tau}_R+
\frac{g}{\sqrt{2}m_W}
\epsilon_{ij}\biggl[\frac{m_{\tau}A_{\tau}}{{\cos}{\beta}}H^i_1\tilde{l}^j_L\tau^*_R\,+
+\, \frac{m_bA_b}{{\cos}{\beta}}H^i_1\tilde{q}^j_L\tilde{b}^*_R-
\frac{m_tA_t}{{\sin}{\beta}}H^i_2\tilde{q}^jt^*_R\biggr] +
\]\vspace*{-6mm}
\[
+ \frac{1}{2}\biggl[M_3\overline{\tilde{g}}\tilde{g}+
M_2\overline{\widetilde{W}^a}\widetilde{W}^a+M_1\overline{\widetilde{B}}\widetilde{B}\biggr]\!,
\]
where $L_{i}$ {and} $Q_{i}$ are a slepton and a squark,
respectively, $SU(2)_{L}$ doublets; $\overline{E}_{j}$ {and}
$(\overline{D}_{j},\overline{U}_{j})$ are a selectron and a squark,
respectively, $SU(2)_{L}$ singlets; and $H_{1}$ {and} $H_{2}$ are
Higgs $SU(2)_{L}$ doublets.

The superpotential $W$ and the potential $V$ depend on more than 100
parameters.\,\,The number of the latter can be reduced to the
following five owing to the theoretical reasoning and the
experimental observations~\cite{12}:\vspace*{-2mm}
\[
m_{0},~ m_{1/2},~ A_{0},~ \tan\beta,~\mathrm{sgn}~\mu,
\]
where $m_{0}$ {and} $m_{1/2}$ are the masses of scalar and
spinor, respectively, superpartners; $A_{0}$ is the parameter of the
trilinear interaction, $\tan\beta$ is the ratio between the vacuum
expectation values of two Higgs doublets, and $\mathrm{sgn}~\mu$ is
the sign of the Higgs mixing parameter.

The application of recent experimental data \cite{13} obtained by
the ATLAS Collaboration for the proton-proton collisions with the
following final states: 1)~0~leptons + 2--6~jets, 2)~0~leptons +
7--10~jets, 3)~0 or 1~lepton + 3~b-quark jets, 4)~1~lepton + jets~+
+~MET (Missing transverse energy), 5)~1 or 2~tau-leptons + jets +
MET, and 6)~2~same-sign leptons~+ +~0 or more than 3 b-quark jets
(see Fig.~1)--allowed us to consider two scenarios of the MSSM model
(they are presented in Table~1) 
\newpage
\begin{center}
\emph{\textbf{Table 1.}} {\it Two scenarios of MSSM model}
\ec
\bec
\begin{tabular}{|c|c|c|c|c|c|}\hline
&$m_{0}$, GeV&$m_{1/2}$, GeV&$A_{0}$, GeV&$\mbox{tan}\beta$&$\mbox{sgn}(\mu)$ \\ \hline
I&1400&800&-2800&30&+1\\
II&5000&600&-10000&30&+1\\ \hline
\end{tabular}
\ec
as such that exceed recent
experimental observations made in six experiments shown in Fig.~1.

\begin{figure}

\centerline{\includegraphics[width=0.8\textwidth]{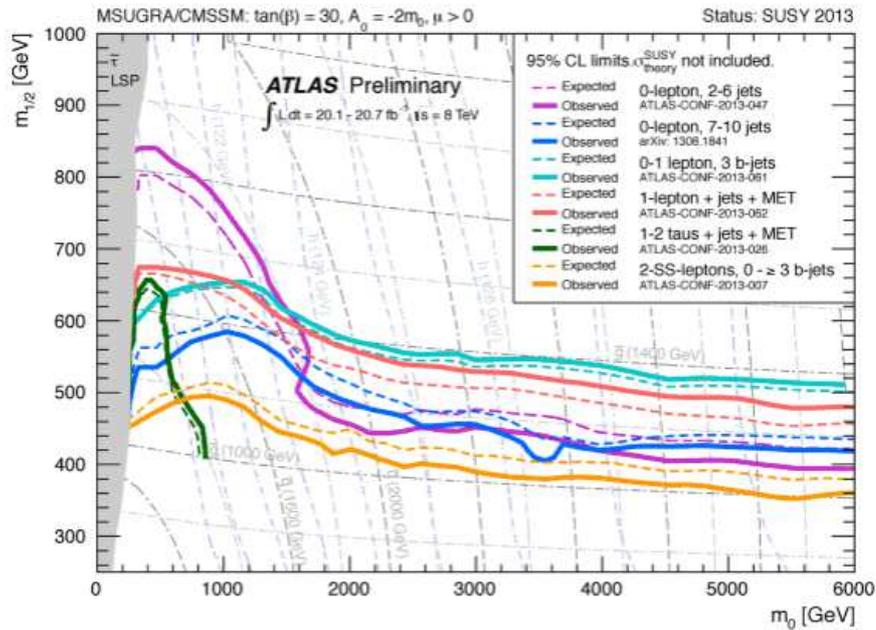}}

\caption{\small Gaugino and scalar masses according to the scale
of the Grand Unification Theory}

\end{figure}

For the calculation of superpartner mass spectra, which are
presented in Table~2, we used the soft\-wa\-re program SOFTSUSY
\cite{14}.\,\,
\newpage
\begin{center}
\emph{\textbf{Table 2.}} {\it Masses of superpartners,
GeV}
\ec
\bec
\begin{tabular}{|c|c|c|c|c|c|c|}\hline
&$m_{\widetilde{u}_{L}}$&$m_{\widetilde{u}_{R}}$&
$m_{\widetilde{d}_{L}}$&$m_{\widetilde{d}_{R}}$&
$m_{\widetilde{g}}$&$m_{\widetilde{\chi}^{0}_{1}}$ \\ \hline
I&2188&2074&2120&2069&1838&344\\
II&5079&5081&5080&5079&1542&273\\ \hline
\end{tabular}
\ec
\vspace*{3mm}
The presented mass spectrum is confined, because the
masses of quark su\-per\-part\-ners are de\-ge\-ne\-ra\-te.\,\,In
Table~2, the masses of first-generation left- and right-chiral
squarks,
(${\widetilde{u}_{L}},{\widetilde{u}_{R}}$) {and} (${\widetilde{d}_{L}%
},{\widetilde{d}_{R}}$), and the masses of glui\-no,
${\widetilde{g}}$, and neut\-ra\-li\-no,
${\widetilde{\chi}_{1}^{0}}$(a candidate for the dark matter), are
re\-pre\-sen\-ted.\,\,The cen\-ter-of-mass energy at the LHC is
planned to achieve a value of
14~TeV, and the luminosity a value of $10^{35}$~$\mathrm{cm}^{-2}%
\mathrm{s}^{-1}$ in 2015. The\-re\-fo\-re, the probability to
ob\-ser\-ve the pro\-ces\-ses of su\-per\-part\-ner production and
subsequent su\-per\-part\-ner decay into quark
jets\,+\,leptons\,+\,MET (e.g., neut\-ra\-li\-no ${\chi}_{1}^{0}$)
depicted in Fig.\,2 will grow considerably.

Using the set of parameters from Table\,\,1, it is pos\-sib\-le to
calculate the cross-sec\-ti\-ons of superpartner production with the
help of the software program PROSPINO \cite{15}.\,\,The
corresponding results, which are listed in Table~3, were obtained
for the transverse cross-sections with regard for the main terms
$\sigma_{\mathrm{LO}}^{\mathrm{PROSPINO}}$ (LO means
\textquotedblleft leading order\textquotedblright) for the
squark-squark, squark-gluino, and glui\-no-glui\-no production, as
well as the corresponding ad\-di\-tio\-nal terms
$\sigma_{\mathrm{NLO}}^{\mathrm{PROSPINO}}$(NLO means
\textquotedblleft next-to-lea\-ding or\-der\textquotedblright),
which result from making allowance for the
re\-nor\-ma\-li\-za\-ti\-on group terms.\,\,The calculations were
carried out for the center-of-mass energy
\mbox{$\sqrt{s}=14~\mathrm{TeV}$.}


\bec
\emph{\textbf{Table 3.}} 
{\it LO and NLO transverse cross-sections (in pb
units) and\\ $K$-factors for superpartners }
\ec
\bec
\begin{tabular}{|c|c|l|l|c|}\hline
&channel&
$\sigma^{\mbox{Prospino}}_{\mbox{LO}}$&
$\sigma^{\mbox{Prospino}}_{\mbox{NLO}}$&
$K^{\mbox{Prospino}}$ \\ \hline
I&squark-squark&0.446E-02&0.488E-02&1.096 \\
&squark-gluino&0.475E-02&0.810E-02&1.704\\ 
&gluino-gluino&0.631E-03&0.204E-02&3.226\\ \hline
II&squark-squark&0.374E-08&0.382E-08&1.022 \\
&squark-gluino&0.169Е-04&0.481Е-04&2.85\\ 
&gluino-gluino&0.595E-02&0.137Е-01&2.307\\ \hline
\end{tabular}
\ec
\vspace*{3mm}
The\, $K$-factor\, is\, the\, ratio\, between\, the\, NLO\, and LO
transverse cross-sections,
$K=\sigma_{\mathrm{NLO}}/\sigma_{\mathrm{LO}}$.\,\,While comparing
the results of calculations for two scenarios, we can mark a
substantial excess of the cross-sections in scenario~I.\,\,This fact
is associated with a large difference between the corresponding
values of parameters $m_{0}$ and $A_{0}$ in both scenarios (see
Table~1).\,\,Attention should also be paid to a large cross-section
of gluino-gluino production in scenario~II, since the mass of such a
gluino amounts to 1542~GeV (Table~2), and the probability of its
production considerably grows with respect to the gluino mass in
scenario~I.\,\,A large value of $K$-factor testifies to the
necessity of more detailed calculations for the cross-sections,
which would include additional (NLO) terms as LO ones in accordance
with the QCD theories.

\begin{figure}

\centerline{\includegraphics{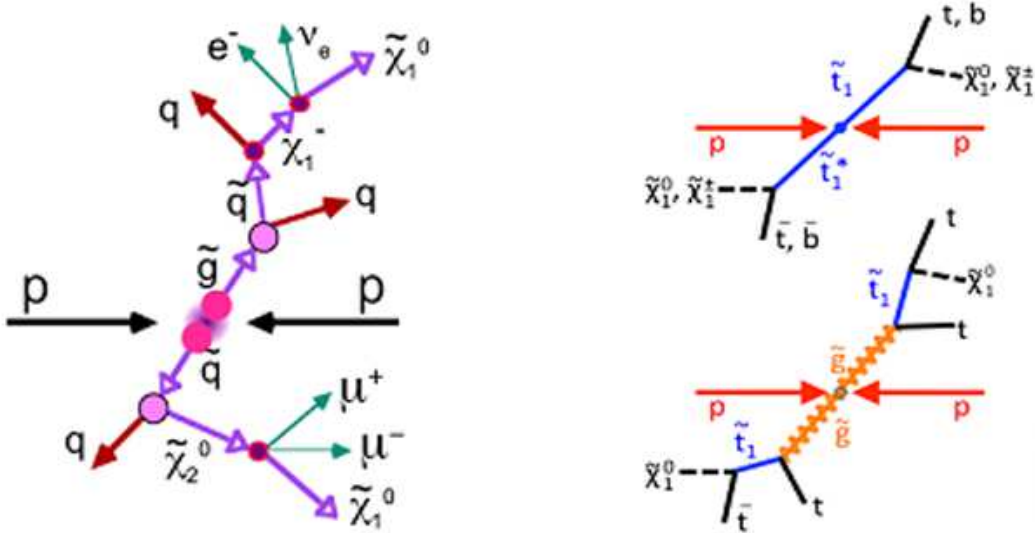}}

\caption{\small Processes of superpartner production at the LHC:
squark-gluino (left panel), squark-squark (right upper panel), and
gluino-gluino (right bottom panel) processes.}

\end{figure}
\section{Conclusions}

Search for the supersymmetry is an important step toward elucidating
the deep contradictions of not only the theoretical, but also
experimental character.\,\,This statement is testified by a number
of recent experimental data, which are difficult to be explained if
disregarding the new physics, including the supersymmetry.\,\,With
the help of the computer simulation (SOFTSUSY, PROSPINO), leading
western scientists \cite{16} carry out a wide range of works in
order to properly respond to the recent experimental data concerning
the search for superpartners.\,\,The permanent updating of
experimental data associated with the processing of large data
arrays in the GRID database brings about the necessity of a new
experimental simulation with the help of computer programs, which
makes our work up-to-date and challenging, because two new scenarios
for the MSSM parameter space are taken into account in accordance
with the last ATLAS experiment.\,\,Since the superpartners are not
so heavy as, e.g., Kaluza--Klein partners or microscopic black
holes, the lower mass limit of which amounts to approximately
5--6~TeV \cite{17}, and the probability of the superpartner
production is high enough, which is confirmed by the results of our
calculations for scenario~I, we hope for that they will be observed
in the nearest future, when the energy and the luminosity at the LHC
will be sufficient for that.

\end{document}